# Two-way Nanoscale automata

Debayan Ganguly[1], Kingshuk Chatterjee[2], Kumar Sankar Ray[3]

[1] Government College of Engineering and Leather Technology, Block-LB, Sector –III, Kolkata-700106

[2] Government College of Engineering and Ceramic technology, Kolkata-700010

[3] Electronics and Communication Sciences Unit, Indian Statistical Institute, Kolkata-700108

debayan3737@gmail.com[1], kingshukchaterjee@gmail.com[2], ksray@isical.ac.in[3]

**Abstract.** In this paper, we show the all final subclass of two-way Watson-Crick automata have the same computational power as the classical two-way Watson-Crick automata. Here we compare the computational power of two-way Watson-Crick automata and two-way Quantum finite automata and we observe that two-way Watson-Crick automata can accept the language L={ww | w∈ {a, b}$^*$}which two way quantum finite automata cannot accept.

## 1      Introduction

The number of transistors in a dense integrated circuit doubles approximately every eighteen months which is known as Moore's law [1]. It is predicted that nanoscale model, will replace the existing classical model in the coming years. Finite state automaton is one of the simplest models of computation for a classical computer. Similarly, quantum finite state automaton and Watson-Crick automata are the simplest nanoscale model for quantum and DNA computing respectively.The basic models of quantum finite state automata are discussed in [2-4]. In search of increasing the language accepting capabilities, different models of 1QFA have been proposed [5-13] by researchers. In 1997, Kondacs and Watrous [3] defined the quantum analogue of 2-way deterministic finite state automata named 2-way quantum finite automata (2QFA). It is more powerful than the 2-way deterministic finite automata. In 2QFA model, the tape head can read the input tape bi-directionally, or it can be stationary. It is more powerful than the classical model because it allows quantum parallelism with the superposition of states on the input tape. Moreover, 2QFA recognizes non-context free languages in linear time, some context-free languages with one-sided error and regular languages.

The first such automata which exploited the DNA property were the Watson-Crick automata [14-15]. Watson-Crick automata are finite automata having two independent heads working on double strands where the characters on the corresponding positions of the two strands are connected by a complementarity relation similar to the Watson-Crick complementarity relation. The movement of the heads although independent of each other is controlled by a single state. Czeizler et.al. [16] introduced the deterministic variants of Watson Crick automata. State complexity of Watson-Crick automata is discussed in [17] and [18]. A two-way Watson-Crick automaton is similar in concept to a two-way finite automaton. The only difference between them is that in two-way Watson-Crick automaton the input tape is double stranded and the content of the second tape is determined in a similar manner as Watson-Crick automaton. The idea of two-way Watson-Crick automata(2NWK) were introduced in [18] but no formal structure is stated.

In this paper, we define subclasses of two-way Watson-Crick automata and show that the all final variant has the same computational power as the classical model. In the last section we show that Watson Crick automata and one-way Watson crick quantum finite state automata accepts the context sensitive language L={ww|w$\in$ {a, b}$^*$} which is not accepted by two-way quantum finite state automata .

## 2  Preliminaries and Definitions

In this Section, we state some important definitions related to the quantum finite automata and Watson-Crick automata.

**Definition 1:** $M = (Q, \Sigma, \delta, q_0, Q_a, Q_r)$ be a 2-way QFA consists of 6-tuple where Q is a finite set of states, $\Sigma$ is an input alphabet, $\delta$ is a transition function, $q_0 \in Q$ is a starting state, $Q_a \in Q$ and $Q_r \in Q$ are sets of accepting and rejecting states.

The automaton halts when it reaches the set of states $Q_a$ or $Q_r$. It continues processing the input when it is in a state which belongs to the set of states $Q_{non}$. The symbols # and $ are used as the left and the right end marker to identify the beginning and ending input word. The working alphabet of M is $\Gamma=\Sigma\cup\{\#,\$\}$.

Trasition function $\delta$ is defined as: $Q \times \Gamma \times Q \times D \to C$, where $\Gamma=\Sigma\cup\{\#,\$\}$ and $D = \{\rightarrow,\uparrow,\rightarrow\}$ represent the left, stationary and right direction of tape head. Transition function must satisfy the following conditions: local probability and orthogonality condition, first separability condition and second separability condition respectively which are shown in [3].

**Definition 2:** Depending on the type of states and transition rules there are two types or subclasses of two-way quantum finite automata. A two-way quantum automaton $M=(Q, \Sigma, \delta, q_0, Q_a, Q_r)$ is

1) stateless(2 NQFA ): If it has only one state, i.e. $Q=F=\{q_0\}$;
2) all-final( 2 FQFA ): If all the states are final, i.e. $Q=F$;

**Definition 3**: The symbol V denotes a finite alphabet.

**Definition 4 :** The set of all finite words over V is denoted by $V^*$, which includes the empty word $\lambda$. The symbol $V^+=V^*- \{\lambda\}$ denotes the set of all non-empty words over the alphabet V.

**Definition 5 :** For $w \in V^*$, the length of w is denoted by |w|.

**Definition 6 :** Let $u \in V^*$ and $v \in V^*$ be two words and if there is some word $x \in V^*$, such that v=ux, then u is a prefix of v, denoted by $u \leq v$.

**Definition 7 :** Two words, u and v are prefix comparable denoted by $u\sim_p v$, if u is a prefix of v or vice versa.

**Definition 8 :** When u is not a prefix of v and v is not a prefix of u then u and v are not prefix comparable which is denoted by $u\not\sim_p v$.

Next we state the important models and definitions associated with Watson-Crick automata.

**Definition 9**: A non-deterministic Watson-Crick automaton is a 6-tuple of the form $M=(V,\rho,Q,q_0,F,\delta)$ where V is an alphabet set, the symbol Q denotes the set of states, the symmetric complementarity relation $\rho \subseteq V \times V$ is called the Watson-Crck complementarity relation, $q_0$ is the initial state and $F \subseteq Q$ is the set of final states. The function $\delta$ contains a finite number of transition rules of the form $q\binom{w_1}{w_2}\to q'$, which denotes that the machine in state q parses $w_1$ in upper strand and $w_2$ in lower strand and goes to state q' where $w_1, w_2 \in V^*$. The symbol $\begin{bmatrix}w_1\\w_2\end{bmatrix}$ is different from $\binom{w_1}{w_2}$. While $\binom{w_1}{w_2}$ is just a pair of strings written in that form instead of $(w_1,w_2)$, the symbol $\begin{bmatrix}w_1\\w_2\end{bmatrix}$ denotes that the two strands are of same length i.e. $|w_1|=|w_2|$ and the corresponding symbols in two strands are complementarity in the sense given by the relation $\rho$. The symbol $\begin{bmatrix}V\\V\end{bmatrix}_\rho = \{\begin{bmatrix}a\\b\end{bmatrix} \mid a, b \in V, (a, b) \in \rho\}$ and $WK_\rho(V)=\begin{bmatrix}V\\V\end{bmatrix}_\rho^*$ denotes the Watson-Crick domain associated with V and $\rho$.

A transition in a Watson-Crick finite automaton can be defined as follows:

For $\binom{x_1}{x_2},\binom{u_1}{u_2},\binom{w_1}{w_2} \in \binom{V^*}{V^*}$ such that $\begin{bmatrix}x_1u_1w_1\\x_2u_2w_2\end{bmatrix} \in WK_\rho(V)$ and q, q' $\in Q$, $\binom{x_1}{x_2}q\binom{u_1}{u_2}\binom{w_1}{w_2} \Rightarrow \binom{x_1}{x_2}\binom{u_1}{u_2}q'\binom{w_1}{w_2}$ iff there is transition rule $q\binom{u_1}{u_2}\to q'$ in $\delta$ and $\Rightarrow^*$ denotes the transitive and reflexive closure of $\Rightarrow$. The language accepted by a Watson-Crick automaton M in the upper strand is $L(M)=\{w_1 \in V^* | q_0\begin{bmatrix}w_1\\w_2\end{bmatrix} \Rightarrow^* \begin{bmatrix}w_1\\w_2\end{bmatrix} q$, with $q \in F$, $w_2 \in V^*, \begin{bmatrix}w_1\\w_2\end{bmatrix} \in WK_\rho(V)\}$.

**Definition 10**: Deterministic Watson-Crick automaton is a Watson-Crick automaton for which if there are two transitions of the form $q\binom{u}{v}\to q'$ and $q\binom{u'}{v'}\to q''$ then $u \not\sim_p u'$ or $v \not\sim_p v'$.

**Definition 11:** A Watson-Crick quantum finite automaton(1WKQFA)[20] is a nine tuple $M=(Q,V,\delta,q_0,Q_{acc},Q_{rej},\rho,\#,\$)$ where Q is a finite set of states, V is the input alphabet, $\delta$ is the transition function, $q_0 \in Q$ is the initial state, $Q_{acc} \subset Q$ and $Q_{rej} \subset Q$ are sets of accepting and rejecting states. The complementarity relation $\rho$ is similar to Watson-Crick complementarity relation. The states in $Q_{acc}$ and $Q_{rej}$ are called halting states and the states in $Q_{non}=Q-(Q_{acc}\cup Q_{rej})$ are called the non-halting states. The symbols '#' and '$' do not belong to V. We use '#' and '$' as the left and right end-markers respectively. The working alphabet of M is $\Gamma=V \cup \{\#,\$\}$.

## 3  Two-way non-deterministic Watson-Crick automata

Two-way non-deterministic Watson-Crick automaton[21] is a 8 tuple, $M=(V,\#,\$,\rho,Q,q_0,F,\delta)$ where V is alphabet, $\#,\$ \notin V$ are the beginning and the end marker respectively. Set of states is denoted by Q, $\rho$ is the symmetric complementarity relation same as that of Watson-Crick automata, $q_0$ is the initial state and $F \subseteq Q$ is the set of final states. $\delta$ contains finite number of transition rules of the form $q\binom{w_1,dir_1}{w_2,dir_2}\rightarrow q'$, which denotes that the machine in state q parses $w_1$ in upper strand in $dir_1$ direction and $w_2$ in lower strand in $dir_2$ direction and goes to state q' where $w_1,w_2 \in V^*\{\#,\$\}^*$ and $dir_1,dir_2 \in \{L,R,0\}$ where L signifies that the head is reading the word in the left direction, R signifies that the head is reading the word in right direction and if a head reads the empty word $\lambda$ it remains in its current position denoted by 0 with the restriction that if $w_1$ or $w_2=V^*\#$ then the corresponding $dir_1$ or $dir_2 \neq L$. The above restriction ensures that the reading heads do not go past the input word on the left side.

**Accepting conditions**

$W_1$ is accepted by M, if starting in state $q_0$ (initial state) with $\begin{bmatrix}\#w_1\$\\\#w_2\$\end{bmatrix}$ and $\begin{bmatrix}w_1\\w_2\end{bmatrix} \in WK_\rho(V)$ on the double stranded input tape and the two heads at the left end of $\#w_1\$$ and $\#w_2\$$. M eventually enters a final state and both the heads fall off the right hand side of the double stranded input tape.

The word $w_1$ is rejected if one of the following 3 conditions occurs:
  i.   The two-way Watson-Crick automaton goes into a loop which is indentified in a similar way as loops in two-way finite automaton.
  ii.  When both the heads fall off the right hand side of the input tape and the machine is in a non-final state.
  iii. If the machine comes to a halt (i.e. there are no transition rules that can be applied for that particular state and inputs the heads are reading) before the heads fall off the right hand side of the input tape.

**Definition 12:** Depending on the type of states and transition rules there are four subclasses of two-way Watson-Crick automata similar to Watson-Crick automata.

A two-way Watson-Crick automaton $M=(V,\#,\$, \rho, Q, q_0, F, \delta)$ is

1) *stateless(2NWK)*: If it has only one state, i.e. $Q=F=\{q_0\}$;

2) *all-final(2FWK)*: If all the states are final, i.e. $Q=F$ ;

3) *simple(2SWK)*: If at each step the automaton reads either from the upper strand or from the lower strand, i.e. for any transition rule $q\binom{w_1,dir_1}{w_2,dir_2}\rightarrow q'$, either $w_1=\lambda$ or $w_2=\lambda$;

4) *1-limlited(21WK)*: If for any transition rule $q\binom{w_1, dir_1}{w_2, dir_2} \rightarrow q'$, we have $|w_1 w_2|=1$.

# 4 Equivalence of Subclasses of two-way Watson-Crick automata

In this Section, we show the equivalence of different subclasses of two-way Watson-Crick automata.

**Theorem 1:** All final two-way Watson-Crick automata have the same computational power as two-way Watson-Crick automata.

**Proof:** Let $M=(V, \#, \$, \rho, Q, q_0, F, \delta)$ be a two-way non-deterministic Watson-Crick automaton. We introduce an all final two-way Watson-Crick automaton $M'=(V, \#, \$, \rho, Q', q_0, \delta')$.

Each transition rule $t$ of the form $q\binom{w_1, dir_1}{w_2, dir_2} \rightarrow q'$ in $\delta$ where $w_1=a_1 a_2 \ldots a_n$ where $|w_1|=n$ and $w_2=b_1 b_2 \ldots b_m$ where $|w_2|=m$ and $a_i, b_j \in V \cup \{\lambda, \#, \$\}$ falls under one of the five classes. The classes are defined as follows:

Class 1: Transition rules of the form $q\binom{w_1, dir_1}{w_2, dir_2} \rightarrow q'$ in $\delta$ where $w_1=a_1 a_2 \ldots a_n$ where $|w_1|=n$ and $w_2=b_1 b_2 \ldots b_m$ where $|w_2|=m$ and $a_n$ and $b_n \neq \$$, i.e. $w_1$ and $w_2$ do not have $\$$ at their ends.

Class 2: Transition rules of the form $q\binom{w_1, R}{w_2, R} \rightarrow q'$ in $\delta$ where $w_1=a_1 a_2 \ldots a_n$ where $|w_1|=n$ and $w_2=b_1 b_2 \ldots b_m$ where $|w_2|=m$, and $a_n, b_n = \$$, i.e. $w_1$ and $w_2$ both have $\$$ at their ends.

Class 3: Transition rules of the form $q\binom{w_1, R}{w_2, dir_2} \rightarrow q'$ in $\delta$ where $w_1=a_1 a_2 \ldots a_n$ where $|w_1|=n$ and $w_2=b_1 b_2 \ldots b_m$ where $|w_2|=m$, $a_n=\$$ and $b_n \neq \$$ i.e. $w_1$ has $\$$ at its end and $w_2$ does not have $\$$ at its end.

Class 4: Transition rules of the form $q\binom{w_1, dir_1}{w_2, R} \rightarrow q'$ in $\delta$ where $w_1=a_1 a_2 \ldots a_n$ where $|w_1|=n$ and $w_2=b_1 b_2 \ldots b_m$ where $|w_2|=m$, $a_n \neq \$$ and $b_n = \$$ i.e. $w_1$ does not have $\$$ at its end and $w_2$ has $\$$ at its end.

Class 5: Either transition rules of the form $q\binom{\$, L}{\lambda, 0} \rightarrow q'$ in $\delta$ or transition rules of the form $q\binom{\$, L}{\$, L} \rightarrow q'$ in $\delta$ or transition rules of the form $q\binom{\lambda, 0}{\$, L} \rightarrow q'$ in $\delta$.

The transition rules of M are modified as follows to form the transition rules of M'

Transition rules of M which fall in class 1 and class 5 are kept same in M'.

For transition rules of M which belong to class 2 two instances can occur;

case 1: For transition $q\begin{pmatrix}w_1,R\\w_2,R\end{pmatrix} \to q'$, where q' is a final state. In this case the transition rules are kept same in M'.

case 2: For transition $q\begin{pmatrix}w_1,R\\w_2,R\end{pmatrix} \to q'$, where q' is a non-final state. In this case the transition rules of M are modified as follows for M'.

For each transition rule $q\begin{pmatrix}w_1,R\\w_2,R\end{pmatrix} \to q'$ in M belonging to class 2 where q' is a non-final state, $q\begin{pmatrix}w_1',R\\w_2',R\end{pmatrix} \to q''$ where $w_1 = w_1'\$$ and $w_2 = w_2'\$$ are introduced in M' and there is no transition from q'' in M'. These new rules in M' ensure that if the heads go off the right end of the tape in M when M is in a non-final state then M' would go to state q'' and would not accept the string as there is no transition from q'' and so the heads of M' will not fall off the right end of the tape. The above stated rules ensure the heads do not fall off the right end of the tape for M' when M does not accept the word. As M' is all final if the heads go off the right end of the tape it will accept the given string.

For transition rules of M which belong to class 3 the following modifications are needed. Class 3 also has two instances similar to class 2.

case 1: For transition $q\begin{pmatrix}w_1,R\\w_2,dir_2\end{pmatrix} \to q'$, where q' is a final state. In this case the transition rules are kept same in M'.

case 2: For transition $q\begin{pmatrix}w_1,R\\w_2,dir_2\end{pmatrix} \to q'$, where q' is a non-final state. In this case the transition rules of M are modified as follows for M'.

For each transition rule $q\begin{pmatrix}w_1,R\\w_2,dir_2\end{pmatrix} \to q'$ in M belonging to class 3 where q' is a non-final state,

$q\begin{pmatrix}w_1',R\\w_2,dir_2\end{pmatrix} \to q'_{u\$}$ where $w_1 = w_1'\$$ is introduced in M' where $q'_{u\$}$ denotes that the head on the upper strand has gone past the right end marker $ in the original machine M on application of the above transition rule.

Only rules having λ on the upper strand are applied to $q'_{u\$}$ because in the actual machine M if the above rules of class 3 are applied then the upper head would have gone past the right end of the tape. So only rules having λ on the upper head can be applied to the original machine M at this stage. As M' replicates M similar thing is done in M' too.



Thus, all the transition rules that can be applied to q' in M with λ on the upper strand and $w_2=a_1a_2\ldots a_n$ and $a_n \neq \$$ in the lower strand can also be applied to q'$_{u\$}$ in M'. Rules having λ on the upper strand and $w_2=a_1a_2\ldots a_n$ and $a_n=\$$ in the lower strand where the transition goes to a final state are applied to q'$_{u\$}$. Finally for rules with λ on the upper strand and $w_2=a_1a_2\ldots a_n$ and $a_n=\$$ in the lower strand where the transition goes to a non-final state, the rules of the form q'$_{u\$}\binom{\lambda,0}{w'_2,R} \to$ q$_{ul\$}$ where $w_2= w_2'\$$ are introduced in M' and there are no transition rules from q$_{ul\$}$ these rules ensure that when M reaches the end of the string on a non-final state then M' goes to q$_{ul\$}$ and M' does not accept the string as there is no transition from q$_{ul\$}$. The above stated rules ensure the heads do not fall off the right end of the tape for M' when heads off M fall off the right end and the state to which M goes is non-final.

Class 4 rules are handled in a similar way to class 3 rules.

All the new states are introduced in Q' along with states in Q.

It is evident from the transition rules introduced in M' that M' accepts the same language as M.

Thus, for every two-way Watson-Crick automaton we obtain an all final two-way Watson-Crick automaton which accepts the same language and every all final two-way Watson-Crick automaton is a two-way Watson-Crick automaton. Therefore, two-way Watson-Crick automaton and all-final two-way Watson-Crick automaton have the same computational power.▫

**Example 1: There is a deterministic two-way Watson-Crick finite automaton that accepts the context sensitive language L={ww|w∈ {a, b}$^*$}.**

**Proof:** M=(V,#,$,ρ,Q,q$_0$,F,δ) is a deterministic two-way Watson-Crick automaton with non-injective complementarity relation ρ that accepts the context sensitive language L={ww|w∈ {a, b}$^*$} where Q={ q$_0$,q$_1$,q$_2$,q$_3$,q$_4$,q$_5$} and F={ q$_5$}.

We define the transitions involved in M as follows:
δ (q$_0$, $\binom{\#}{\#}$, 1,1)= q$_0$ , δ (q$_0$, $\binom{x}{xx}$, 1,1)= q$_0$ , δ (q$_0$, $\binom{\lambda}{\$}$, 0, −1)= q$_1$ ,

δ (q$_1$, $\binom{\lambda}{x}$, 0, −1)= q$_1$ , δ (q$_1$, $\binom{\lambda}{\$}$, 0,1)= q$_2$ , δ (q$_2$, $\binom{x}{x}$, 1,1)= q$_2$ , δ (q$_2$, $\binom{\$}{\lambda}$, −1,0)= q$_3$ , δ (q$_3$, $\binom{\lambda}{x}$, 0,1)= q$_3$ , δ (q$_3$, $\binom{\lambda}{x}$, 0, −1)= q$_4$ , δ (q$_4$, $\binom{\$}{\$}$, 1,1)= q$_5$

The above automata works in the following manner after reading the first end marker #.The upper head reads a symbol and the lower head reads two symbols. Now if the length of the input word |w| is odd then the lower head will never reach $ because it is reading two symbols at a time. Thus it will terminate in a non final state. Thus all odd strings are eliminated. If the input string is even, then the upper head stops at the first middle element (as it is even length string, there are two middle element).Now the

lower head is moved to the beginning of the string and the first element is matched with the ($1^{st}$ middle element + 1)$^{th}$ element and this matching continues until $ is reached on upper head. If there is a mismatch somewhere then no transition is defined thus automaton rejects the input. The upper head will reach $ only if the two halves match i.e. the input is of the form ww. Then the lower head also moves to $ to accept the input. Then the automation only accepts string of the form ww.

**Theorem 2: $L_{2NWK}$-$L_{2QFA} \neq \emptyset$, where $L_{2NWK}$ is the set of all languages accepted by two-way non-deterministic Watson-Crick automata and $L_{2QFA}$ is the set of all languages accepted by two- way quantum automata.**

**Proof:** From Example 1, we know that there is a two-way Watson-Crick automaton that accepts the context sensitive language L={ww|w∈ {a, b}$^*$} and from paper[22] we know that the context sensitive language L={ww|w∈ {a, b}$^*$} is not accepted by two-way quantum automaton which proves the above Theorem.

**Theorem 3: $L_{1WKQFA}$-$L_{2QFA} \neq \emptyset$, where $L_{1WKQFA}$ is the set of all languages accepted by Watson-Crick quantum finite automata and $L_{2QFA}$ is the set of all languages accepted by two-way quantum automata.**

**Proof:** From [20], we know that there is a one-way Watson-Crick quantum finite automaton that accepts the context sensitive language L={ww|w∈ {a, b}$^*$} and from paper[22]we know that the context sensitive language L={ww|w∈ {a, b}$^*$} is not accepted by two-way quantum automaton which proves the above Theorem.

## 5   Conclusion

We investigated the computational power of different subclasses of two-way non-deterministic Watson-Crick automata. We show that all final two-way Watson-Crick automata and two-way Watson-Crick automata have the same computational power. In the last section we showed that Watson Crick automata and 1-way Watson crick quantum finite state automata accepts a language L={ww|w∈ {a, b}$^*$} which is not accepted by two-way quantum finite state automata.